\documentclass[english]{lni}

\usepackage{tikz}
\usetikzlibrary{positioning, arrows.meta, shapes.geometric, calc, shadows, decorations.pathmorphing}
\usepackage{booktabs}
\usepackage{graphicx}

\newcommand{\ToolEval}{JudgeGPT}
\newcommand{\ToolGen}{RogueGPT}
\newcommand{\ToolRepoGen}{github.com/aloth/RogueGPT}
\newcommand{\ToolRepoEval}{github.com/aloth/JudgeGPT}

\newcommand{\selfcite}{\cite{loth2024blessing}}

\usepackage{array}
\begin{document}

\title[Interrupting the Chain]{Interrupting the Chain: Human Perception of AI-Generated Disinformation Through a Kill Chain Lens}

\author[1]{Alexander Loth}{alexander.loth@stud.fra-uas.de}{0009-0003-9327-6865}
\author[1]{Martin Kappes}{kappes@fra-uas.de}{0000-0002-8768-8359}
\author[2]{Marc-Oliver Pahl}{marc-oliver.pahl@imt-atlantique.fr}{0000-0001-5241-3809}
\affil[1]{Frankfurt University of Applied Sciences\\Nibelungenplatz 1\\60318 Frankfurt am Main\\Germany}
\affil[2]{IMT Atlantique, UMR IRISA, Chaire Cyber CNI\\Brest\\France}

\maketitle

\begin{abstract}
Generative AI enables customized misinformation at scale, yet defenses remain largely reactive. We present empirical findings from a human-subject study ($n{=}504$ participants, $n{=}2{,}438$ judgments) in which users classified news fragments by origin (human vs.\ machine) and veracity (real vs.\ fake). We organize results using an adapted cybersecurity kill chain as a \emph{taxonomy for intervention}, mapping perception data onto stages of a cognitive attack lifecycle. Three key findings emerge: (1)~a \emph{perception-accuracy gap} where heightened suspicion does not improve detection; (2)~modern LLMs frequently produce human-indistinguishable text; and (3)~an \emph{asymmetric cognitive fatigue} effect where fake-news detection degrades by 10.2 percentage points under sustained exposure while AI-origin detection remains stable. These findings identify candidate intervention points for proactive defense against AI-driven disinformation.
\end{abstract}

\section{Introduction}

Generative AI has transformed disinformation from isolated incidents into systematic cognitive attacks that threaten trust in online information ecosystems~\cite{loth2024blessing}. Large language models enable malicious actors to produce culturally nuanced, plausible narratives at scale---a process we have elsewhere termed \emph{industrialized deception}~\cite{loth2026collateraleffects}, and refer to throughout this paper as \emph{AI-generated misinformation}---creating a fundamental asymmetry: falsehoods are produced faster than they can be debunked~\cite{ferrara2024genai,chen2023combating}.

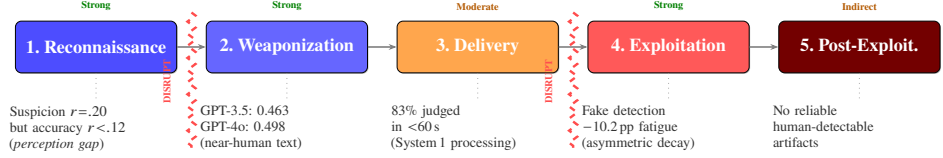
\begin{figure}[t]
\centering
\resizebox{\linewidth}{!}{%
\begin{tikzpicture}[
    stage/.style={
        draw, rounded corners=3pt, minimum width=2.8cm, minimum height=0.9cm,
        font=\small\bfseries, text=white, align=center, drop shadow={shadow xshift=0.5pt, shadow yshift=-0.5pt}
    },
    finding/.style={
        font=\scriptsize, align=left, text width=3.0cm, text=black!80
    },
    defense/.style={
        font=\scriptsize\itshape, align=left, text width=3.0cm, text=black!55
    },
    arr/.style={-{Stealth[length=4pt]}, thick, draw=black!50},
    interrupt/.style={dashed, very thick, red!70, decorate, decoration={zigzag, segment length=6pt, amplitude=2pt}}
]

\node[stage, fill=blue!70] (s1) {1. Reconnaissance};
\node[stage, fill=blue!60, right=0.5cm of s1] (s2) {2. Weaponization};
\node[stage, fill=orange!70, right=0.5cm of s2] (s3) {3. Delivery};
\node[stage, fill=red!65, right=0.5cm of s3] (s4) {4. Exploitation};
\node[stage, fill=red!45!black, right=0.5cm of s4] (s5) {5. Post-Exploit.};

\draw[arr] (s1) -- (s2);
\draw[arr] (s2) -- (s3);
\draw[arr] (s3) -- (s4);
\draw[arr] (s4) -- (s5);

\node[finding, below=0.45cm of s1] (f1) {Suspicion $r{=}.20$\\but accuracy $r{<}.12$\\(\emph{perception gap})};
\node[finding, below=0.45cm of s2] (f2) {GPT-3.5: 0.463\\GPT-4o: 0.498\\(near-human text)};
\node[finding, below=0.45cm of s3] (f3) {83\% judged\\in $<$60\,s\\(System\,1 processing)};
\node[finding, below=0.45cm of s4] (f4) {Fake detection\\$-$10.2\,pp fatigue\\(asymmetric decay)};
\node[finding, below=0.45cm of s5] (f5) {No reliable\\human-detectable\\artifacts};

\draw[dotted, black!40] (s1.south) -- (f1.north);
\draw[dotted, black!40] (s2.south) -- (f2.north);
\draw[dotted, black!40] (s3.south) -- (f3.north);
\draw[dotted, black!40] (s4.south) -- (f4.north);
\draw[dotted, black!40] (s5.south) -- (f5.north);

\node[font=\tiny, text=green!50!black, above=0.08cm of s1] {\textbf{Strong}};
\node[font=\tiny, text=green!50!black, above=0.08cm of s2] {\textbf{Strong}};
\node[font=\tiny, text=orange!70!black, above=0.08cm of s3] {\textbf{Moderate}};
\node[font=\tiny, text=green!50!black, above=0.08cm of s4] {\textbf{Strong}};
\node[font=\tiny, text=orange!70!black, above=0.08cm of s5] {\textbf{Indirect}};

\draw[interrupt] ($(s1.east)!0.5!(s2.west)+(0,0.6)$) -- ($(s1.east)!0.5!(s2.west)+(0,-1.8)$);
\draw[interrupt] ($(s3.east)!0.5!(s4.west)+(0,0.6)$) -- ($(s3.east)!0.5!(s4.west)+(0,-1.8)$);

\node[font=\tiny\bfseries, text=red!70, rotate=90, anchor=south] at ($(s1.east)!0.5!(s2.west)+(-0.25,-0.6)$) {DISRUPT};
\node[font=\tiny\bfseries, text=red!70, rotate=90, anchor=south] at ($(s3.east)!0.5!(s4.west)+(-0.25,-0.6)$) {DISRUPT};

\end{tikzpicture}
}%
\caption{The Disinformation Kill Chain as a taxonomy for intervention. Each stage is annotated with key empirical findings from our study. Evidence strength is indicated above each stage. Red zigzag lines mark candidate disruption points where defenses can interrupt the attack lifecycle. Stages 1, 2, and 4 have strong empirical grounding; Stages 3 and 5 rely on indirect evidence.}
\label{fig:killchain}
\end{figure}

Current defenses---fact-checking, content moderation, debunking---operate \emph{after} cognitive exploitation has occurred. A proactive framework is needed that identifies intervention points \emph{before} impact. We adapt the cybersecurity kill chain concept~\cite{nimmo2023phase} as a \emph{taxonomy for intervention}: a conceptual scaffold that organizes empirical signals into actionable defense layers, without claiming that disinformation campaigns literally proceed through discrete sequential phases. Our five-stage model (Figure~\ref{fig:killchain}) comprises \textbf{Reconnaissance} (profiling targets), \textbf{Weaponization} (creating AI content), \textbf{Delivery} (dissemination), \textbf{Exploitation} (cognitive effects), and \textbf{Post-Exploitation} (evading attribution).

\textbf{Contribution.} We present a quantitative assessment of human perception data ($n{=}504$ participants, $n{=}2{,}438$ judgments) collected via \ToolEval{}~\selfcite{} and organize findings through this kill chain lens to characterize stage-specific vulnerabilities and candidate intervention points. This work complements our prior analyses of causal susceptibility factors~\cite{loth2026eroding}, expert perceptions of the verification crisis~\cite{loth2026verification,loth2026aiesexperts,loth2026hksexperts}, dual-axis human perception of LLM-generated news~\cite{loth2026judgegptwebsci}, the cognitive indistinguishability threshold~\cite{loth2026symposium}, and cryptographic provenance countermeasures~\cite{loth2026originlenswebsci}.

\section{Related Work}

Kill chain frameworks have been proposed for disinformation~\cite{nimmo2023phase,Sedova2021,Schneier2019}, and human AI-text detection has been studied in isolation~\cite{Clark2021,Jakesch2022}. Existing conceptual models of influence operations differ in granularity and intent: Nimmo and Hutchins' ten-phase kill chain~\cite{nimmo2023phase} runs from acquiring assets to enabling longevity; the DISARM/AMITT framework and Sedova et al.~\cite{Sedova2021} decompose operations into tactics and techniques analogous to MITRE ATT\&CK; and Schneier~\cite{Schneier2019} frames information operations as attacks on democratic cognition. These frameworks primarily describe \emph{operator} behavior. Our contribution is complementary and distinct: we retain a compact five-stage lifecycle but anchor each stage in \emph{measured human-perception data}, shifting the unit of analysis from what attackers do to where human targets are cognitively vulnerable.

\textbf{Why a kill chain, and where the analogy holds.} The cybersecurity kill chain is useful here because it makes \emph{intervention points} explicit: a defender does not need to block every stage, only to interrupt one. The analogy holds for the intervention logic (stage-specific, layered defense) and for the notion that early disruption is cheaper than late remediation. It holds \emph{less well} as a causal or strictly sequential process model: disinformation stages overlap, iterate, and run in parallel, and a target may be exploited without a preceding tailored reconnaissance step. We therefore use the kill chain explicitly as a \emph{taxonomy for intervention}, not as a claim that campaigns proceed through discrete, ordered phases.

However, prior kill-chain treatments of disinformation---whether framed around campaign spread, operator TTPs, or, more recently, narrative-sequencing of AI-generated content~\cite{eunomia2023killchain,sensity2025narrative}---remain grounded in \emph{operator} behavior or content analysis. To our knowledge, none has mapped large-scale human \emph{perception} data onto the attack lifecycle to identify stage-specific cognitive vulnerabilities. Our work bridges these threads by providing empirical grounding for a kill chain taxonomy using real perception data. While validated instruments like the Misinformation Susceptibility Test (MIST)~\cite{maertens2024mist} measure veracity judgment at the headline level, \ToolEval{} adds a second dimension---\emph{origin} (human vs.\ machine)---and evaluates full-text fragments, capturing richer cues relevant to the generative AI era.

\section{Methodology}

\ToolGen{} generates news fragments using seven LLMs (GPT-3.5-turbo, GPT-4, GPT-4o, Llama-2-13b, Gemma-7b, Phi-3-mini, Mistral-7b) with systematic prompt variation across four languages (English, French, German, Spanish), five styles, and three formats. Human-written content from legitimate outlets and verified fake-news databases supplements the corpus. The generation pipeline and its implications for industrialized disinformation are detailed in~\cite{loth2026collateraleffects}.

\ToolEval{} collected evaluations from $n{=}504$ participants who provided demographics and then evaluated randomly sampled fragments using three 0--1 sliders: \emph{HumanMachineScore} (origin), \emph{LegitFakeScore} (veracity), and \emph{TopicKnowledgeScore} (familiarity). Accuracy was coded as correct when $\text{Score} \geq 0.5$ matched ground truth. Analysis comprised Pearson correlations, independent t-tests, temporal fatigue analysis, and LLM performance comparison. A causal analysis of the susceptibility factors is presented in~\cite{loth2026eroding}.

\section{Findings Through the Kill Chain Lens}

We organize our key findings by kill chain stage. Table~\ref{tab:killchain_summary} provides an overview. We focus on the four stages with empirical grounding; Post-Exploitation is discussed briefly as it relies on indirect evidence.

\begin{table}[t]
\centering
\caption{Mapping Key \ToolEval{} Findings to the Disinformation Kill Chain}
\label{tab:killchain_summary}
\small
\setlength{\tabcolsep}{3pt}\begin{tabular}{>{\raggedright\arraybackslash\hyphenpenalty=10000}p{0.18\linewidth} >{\raggedright\arraybackslash\hyphenpenalty=10000}p{0.38\linewidth} >{\raggedright\arraybackslash\hyphenpenalty=10000}p{0.38\linewidth}}
\toprule
\textbf{Kill Chain Stage} & \textbf{Empirical Finding} & \textbf{Defensive Implication} \\
\midrule
\textbf{Reconnaissance} & Fake-news familiarity correlates with higher suspicion ($r{=}0.20$) but \emph{not} accuracy ($r{<}0.12$). Overall: 58.4\% AI detection, 68.1\% fake detection. & Perception-accuracy gap is exploitable: both low-literacy and overconfident populations are vulnerable through different vectors. \\
\addlinespace
\textbf{Weaponization} & GPT-3.5-turbo (0.463) and GPT-4o (0.498) produce text frequently perceived as human-written. AI-generated \emph{legitimate} content: origin detection 44.6\% (below chance). & Human detection of state-of-the-art AI text is unreliable; technical provenance solutions~\cite{loth2026originlenswebsci} are needed. \\
\addlinespace
\textbf{Delivery} & Fast judgments (${\leq}$34s): origin accuracy 47.8\% (below chance), veracity 67.3\%. Deliberate ($>{34}$s): 56.4\% and 74.1\%. & System~1 processing is exploitable; friction-based interventions can shift users to deliberative evaluation. \\
\addlinespace
\textbf{Exploitation} & Fake detection accuracy degrades 10.2pp (71.2\%$\to$60.9\%) under sustained exposure; AI detection stable at ${\sim}$56\%. & ``Flooding the zone'' selectively erodes content evaluation; pacing interventions and cognitive scaffolding are needed. \\
\bottomrule
\end{tabular}
\end{table}

\subsection{Reconnaissance: The Perception-Accuracy Gap}

Analysis of demographic correlations reveals that participants with higher fake-news familiarity exhibit moderately higher suspicion scores ($r{=}0.20$ for HumanMachineScore), yet accuracy-based analysis shows negligible correlations ($|r|{<}0.12$). Overall detection accuracy is modest: 58.4\% for AI content (near the 50\% chance baseline) and 68.1\% for fake content. This \emph{perception-accuracy gap}---where heightened suspicion does not translate to improved detection---represents a distinct cognitive vulnerability consistent with the ``truth-default'' erosion we analyze causally in~\cite{loth2026eroding}. We note this as an observed association rather than a tested attack: adversaries \emph{may} be able to exploit not only low-literacy populations but also those whose inflated confidence leads to less careful evaluation, though our data does not directly test adversarial exploitation.

\subsection{Weaponization: LLM Output Quality}

Table~\ref{tab:model_performance} shows the average HumanMachineScore by generating model. Modern commercial LLMs (GPT-3.5-turbo: 0.463; GPT-4o: 0.498) frequently produce text perceived as human-written, while smaller open-source models are somewhat more detectable (0.519--0.545). The overlapping confidence intervals and proximity to the 0.50 midpoint confirm that human detection of high-quality AI text is unreliable at the individual judgment level.

\begin{table}[h]
\centering
\small
\caption{HumanMachineScore by LLM (lower = more human-like).}
\label{tab:model_performance}
\begin{tabular}{lccr}
\toprule
\textbf{Model} & \textbf{N} & \textbf{Score} & \textbf{95\% CI} \\
\midrule
\texttt{gpt-35-turbo} & 324 & 0.463 & [0.430, 0.495] \\
\texttt{gpt-4o} & 343 & 0.498 & [0.464, 0.531] \\
\texttt{mistral-7b} & 477 & 0.519 & [0.489, 0.548] \\
\texttt{llama-2-13b} & 532 & 0.531 & [0.505, 0.558] \\
\texttt{Phi-3-mini} & 147 & 0.543 & [0.501, 0.585] \\
\texttt{gemma-7b} & 532 & 0.545 & [0.517, 0.572] \\
\bottomrule
\end{tabular}
\end{table}

A critical nuance emerges when analyzing detection accuracy by content type (Figure~\ref{fig:accuracy_type_speed}a). For AI-generated \emph{legitimate} content, origin detection drops to 44.6\%---\emph{below} the 50\% chance baseline---meaning participants more often misidentified truthful AI text as human-written. In contrast, AI-generated \emph{fake} content is somewhat more detectable (58.6\% origin accuracy). Veracity detection follows a similar pattern: 65.8\% for AI-legitimate vs.\ 75.1\% for AI-fake content. Interpreted cautiously, this indicates that AI-generated truthful content is particularly \emph{difficult to identify as synthetic}; we do not claim it is more harmful in practice, but its low detectability could make it an effective vehicle for building source credibility ahead of later fabrications.

\subsection{Delivery: Speed-Accuracy Trade-off}

\begin{figure}[h]
\centering
\includegraphics[width=\linewidth]{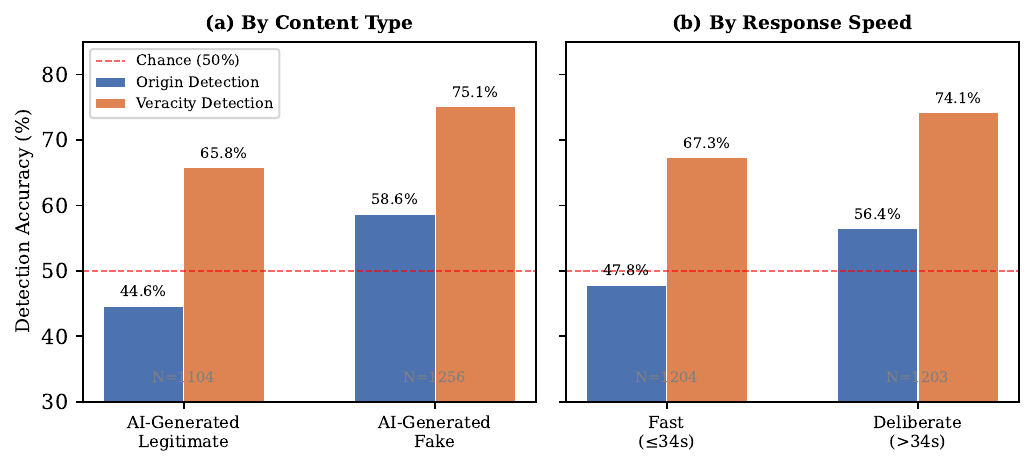}
\caption{Detection accuracy by (a)~content type and (b)~response speed. AI-generated legitimate content falls below chance for origin detection (44.6\%). Fast judgments (${\leq}$34s) also fall below chance, showing System~1 processing undermines detection.}
\label{fig:accuracy_type_speed}
\end{figure}

Response-time analysis reveals that 83\% of judgments were made in under 60 seconds (median: 34s), consistent with heuristic-driven ``System~1'' processing~\cite{kahneman2011thinking}. A median-split analysis (Figure~\ref{fig:accuracy_type_speed}b) shows that deliberate responses (${>}$34s) yield higher accuracy on both dimensions: origin detection improves from 47.8\% to 56.4\%, and veracity detection from 67.3\% to 74.1\%. We report these as \emph{descriptive} differences from a median split; because responses are nested within participants, we did not conduct a repeated-measures test here and treat the effect as exploratory pending mixed-effects modeling (see Future Work). Fast responses fall \emph{below} chance for origin detection, consistent with the interpretation that rapid, shallow processing undermines the ability to identify AI-generated content. This motivates---rather than proves the efficacy of---interventions that introduce ``friction'' at the point of consumption, such as interstitial prompts or brief mandatory dwell times, intended to shift users from System~1 toward System~2 processing.

\subsection{Exploitation: Asymmetric Cognitive Fatigue}

\begin{figure}[h]
\centering
\includegraphics[width=\linewidth]{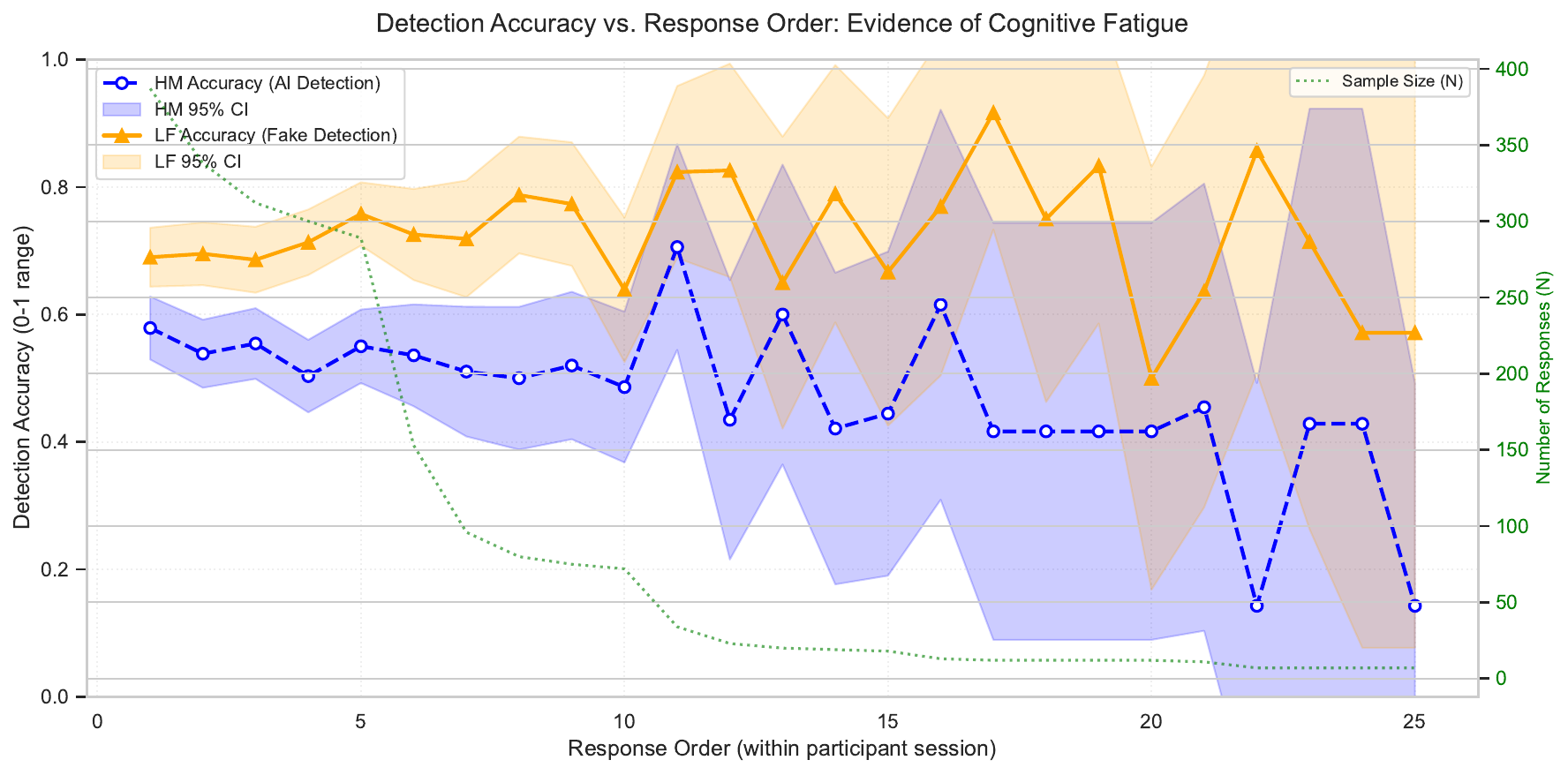}
\caption{Asymmetric cognitive fatigue: fake detection accuracy declines 10.2pp while AI detection remains stable at ${\sim}$56\%.}
\label{fig:learning_fatigue}
\end{figure}

The most striking finding concerns temporal dynamics. As participants evaluate more fragments, fake-news detection accuracy declines from 71.2\% (responses 1--10) to 60.9\% (responses 21--30), a 10.2 percentage point degradation (Figure~\ref{fig:learning_fatigue}). Crucially, AI-origin detection remains stable at approximately 56\% throughout. We report this asymmetry as a \emph{descriptive, exploratory} trend based on binned response positions aggregated across participants; a confirmatory repeated-measures or mixed-effects analysis that accounts for per-participant variation and dropout is left to future work. Read cautiously, the pattern suggests that content-quality evaluation may be more cognitively demanding than origin attribution and may degrade selectively under sustained information load. If this holds, high-volume campaigns could function not merely as amplification but as pressure on cognitive endurance---plausibly exhausting the evaluative capacity that matters most---though our study does not directly test adversarial volume strategies.

\subsection{Post-Exploitation}

The near-midpoint scores of high-quality LLMs (Table~\ref{tab:model_performance}) suggest that AI-generated text inherently lacks consistent human-detectable artifacts, providing built-in plausible deniability---a challenge that motivates cryptographic provenance approaches such as~\cite{loth2026originlenswebsci}. This stage requires stronger empirical grounding in future work, such as longitudinal belief-persistence measurements.

\section{Discussion and Implications}

Our findings support three takeaways for research on AI-generated disinformation. We first restate the empirical observations, then our interpretation. \emph{Empirically}, we observe (i) a perception-accuracy gap, (ii) below-chance origin detection for high-quality and truthful AI text, and (iii) a descriptive asymmetric decline in fake-news detection under sustained exposure. \emph{Interpreting} these through the kill chain: first, the perception-accuracy gap challenges the assumption that awareness campaigns alone improve resilience---metacognitive calibration may be as important as factual inoculation. Second, the asymmetric fatigue pattern is consistent with volume-based attacks being strategically rational, targeting the cognitive process (content evaluation) that degrades under load while leaving origin attribution unaffected; we frame this as a hypothesis motivated by our data rather than a demonstrated attack. Third, the kill chain taxonomy, while not a causal model, provides a useful organizational framework for mapping heterogeneous empirical signals onto distinct defensive opportunities. An expert survey of disinformation specialists corroborates these concerns, highlighting systemic risks of ``epistemic fragmentation'' and skepticism toward purely technical detection~\cite{loth2026verification}.

\textbf{Who can intervene, and how.} Table~\ref{tab:killchain_summary} maps findings to stages; here we make the interventions concrete and assign them to stakeholders, while keeping claims modest.
\begin{itemize}
\item \emph{Platform operators} can add \textbf{friction} at Delivery---lightweight interstitials or brief dwell-time prompts before resharing---to nudge users from fast System~1 toward deliberate System~2 evaluation, and can \textbf{pace} exposure to counter fatigue (e.g., rate-limiting dense misinformation feeds).
\item \emph{AI developers and standards bodies} can strengthen \textbf{provenance} at Weaponization/Post-Exploitation---machine-readable content credentials (e.g., C2PA) and watermarking---so that low human detectability of synthetic text is backstopped by verifiable signals~\cite{loth2026originlenswebsci}. Domain-level credibility priors such as \emph{CRED-1}~\cite{loth2026cred1} offer a complementary, on-device signal for pre-bunking at the Delivery stage.
\item \emph{Educators} can target the perception-accuracy gap at Reconnaissance with \textbf{cognitive scaffolding}: calibration exercises and prebunking that teach \emph{when} to distrust one's own confidence, not only \emph{what} is false.
\item \emph{Policymakers} can require provenance disclosure and support independent evaluation, complementing platform- and education-level measures.
\end{itemize}
We present these as plausible, testable directions rather than validated countermeasures; each requires dedicated evaluation.

\textbf{Limitations.} The study was conducted in a controlled environment lacking the social dynamics of real platforms (limited ecological validity); reshare pressure, social endorsement cues, and algorithmic amplification are absent. The sample is primarily European/North American, and although \ToolGen{} produces multilingual stimuli, cross-language perception effects were not analyzed in this short paper. Content was text-only, excluding multimodal (image/video/audio) disinformation and social-network propagation, both of which likely alter detectability. Findings are tied to the specific foundation models tested; rapidly evolving models may shift the reported detection rates, so absolute numbers should be read as a snapshot. Finally, the kill chain mapping is interpretive rather than experimentally validated per stage, and the speed and fatigue results are descriptive/exploratory pending repeated-measures analysis.

\textbf{Data availability.} The \ToolEval{} perception dataset (DOI: \texttt{10.5281/zenodo.18703385}) and \ToolGen{} stimulus corpus (DOI: \texttt{10.5281/zenodo.18703138}) are archived on Zenodo under restricted access with a data use agreement. Complementary datasets from our broader research program are likewise openly archived: the expert-survey data analyzed in~\cite{loth2026aiesexperts,loth2026hksexperts} (DOI: \texttt{10.5281/zenodo.18703601}) and the \emph{CRED-1} domain-credibility dataset~\cite{loth2026cred1} (DOI: \texttt{10.5281/zenodo.18769460}). Source code is available at \texttt{\ToolRepoGen} (\ToolGen{}) and \texttt{\ToolRepoEval} (\ToolEval{}).

\textbf{Future work} should strengthen the framework-data linkage through integrative modeling (e.g., mixed-effects and multivariate regression across stages that account for per-participant nesting), add cross-language analysis, compare human performance against automated detectors, extend the framework to \emph{multimodal} content (image, video, audio) and \emph{social-network propagation} dynamics, and empirically test the Delivery and Post-Exploitation stages through ecological and longitudinal designs.

\section{Conclusion}

This paper presented empirical evidence from a human-subject study ($n{=}504$, $n{=}2{,}438$ judgments) organized through an adapted cybersecurity kill chain. The perception-accuracy gap, LLM weapon efficacy, and asymmetric cognitive fatigue effect collectively argue for a shift from reactive fact-checking to proactive, stage-specific defense. Understanding where humans become vulnerable is as important as understanding how AI generates misinformation: by mapping perception onto the attack lifecycle, we can identify where---and in whom---to interrupt it. As generative models continue to erase the artifacts that once betrayed synthetic text, the decisive line of defense shifts from detecting the machine to fortifying the mind.

\printbibliography

\end{document}